\documentclass[12pt]{iopart}

\usepackage{iopams}
\usepackage{graphicx}

\begin{document}

\title[]{Cold nuclear effects on heavy flavours (a review)}

\author{R. Granier de Cassagnac}

\address{Laboratoire Leprince-Ringuet, \'Ecole polytechnique/IN2P3, Palaiseau, 91128 France}
\ead{raphael@in2p3.fr}

\begin{abstract}
Before wondering about the quark-gluon plasma (QGP), one has to take into account various cold (normal) nuclear matter effects, that can be probed through p+A like collisions. This article aims at reviewing the current results (and understanding) of these effects on heavy quarks and quarkonia production.\end{abstract}

%\maketitle

\section{Foreword}

The understanding of normal nuclear effects on heavy flavours production recently became crucial to understand the $J/\psi$ nuclear modification factors measured with the PHENIX experiment~\cite{Adare:2006ns}. In particular, the observed rapidity dependence of $J/\psi$ suppression is reversed with respect to what one would naively expect from density-induced suppression mechanisms, such as the colour screening originally proposed as a QGP signature~\cite{Matsui:1986dk}. $J/\psi$ are more suppressed at higher rapidity, where the density is lower. The question is: Could cold nuclear matter effects explain this behaviour? Or do we need to invoke the coalescence of $c$ and $\bar{c}$ quarks coming from uncorrelated pairs instead? This question is so central that the two contributions on quarkonia suppression in heavy ion collisions we heard at this conference discussed it in details~\cite{LindenLevy:SQM08,Gunji:SQM08}. In this paper, I focus on cold nuclear matter only, on most of the available p+A like collisions (from SPS, FNAL, HERA-B and RHIC) and on all the heavy flavour observables ($J/\psi$, $\psi'$, $\Upsilon$ and open charm...). In the following sections, I show the various dependencies of the heavy flavours productions, as they appeared chronologically.

But before to look at data, one needs to define a somewhat arbitrary boundary between what we call "hot" and "cold". Initial state effects (EMC effect, parton antishadowing, shadowing or saturation, energy loss and Cronin effect, intrinsic charm...) obviously belong to the cold sector. Among the final state effects, absorption by incoming nucleons is usually classified as a cold effect too. On the contrary, absorption by comovers (of often unknown partonic or hadronic nature) is an effect of the hot matter. Obviously, so are the long awaited colour screening and quark recombination or coalescence. In a short and pragmatic way, I define as cold what can be grasped in p+A like collisions, while hot is what happens in addition when one looks at A+A collisions. In this paper, we shall then mostly look at p+A like collisions.

% Caution about saturation ?

\section{At SPS, some charmonia and a nuclear length scaling}

\begin{figure}[htb]
\begin{center}
  \begin{tabular}{cc}
  \includegraphics[width=8.cm]{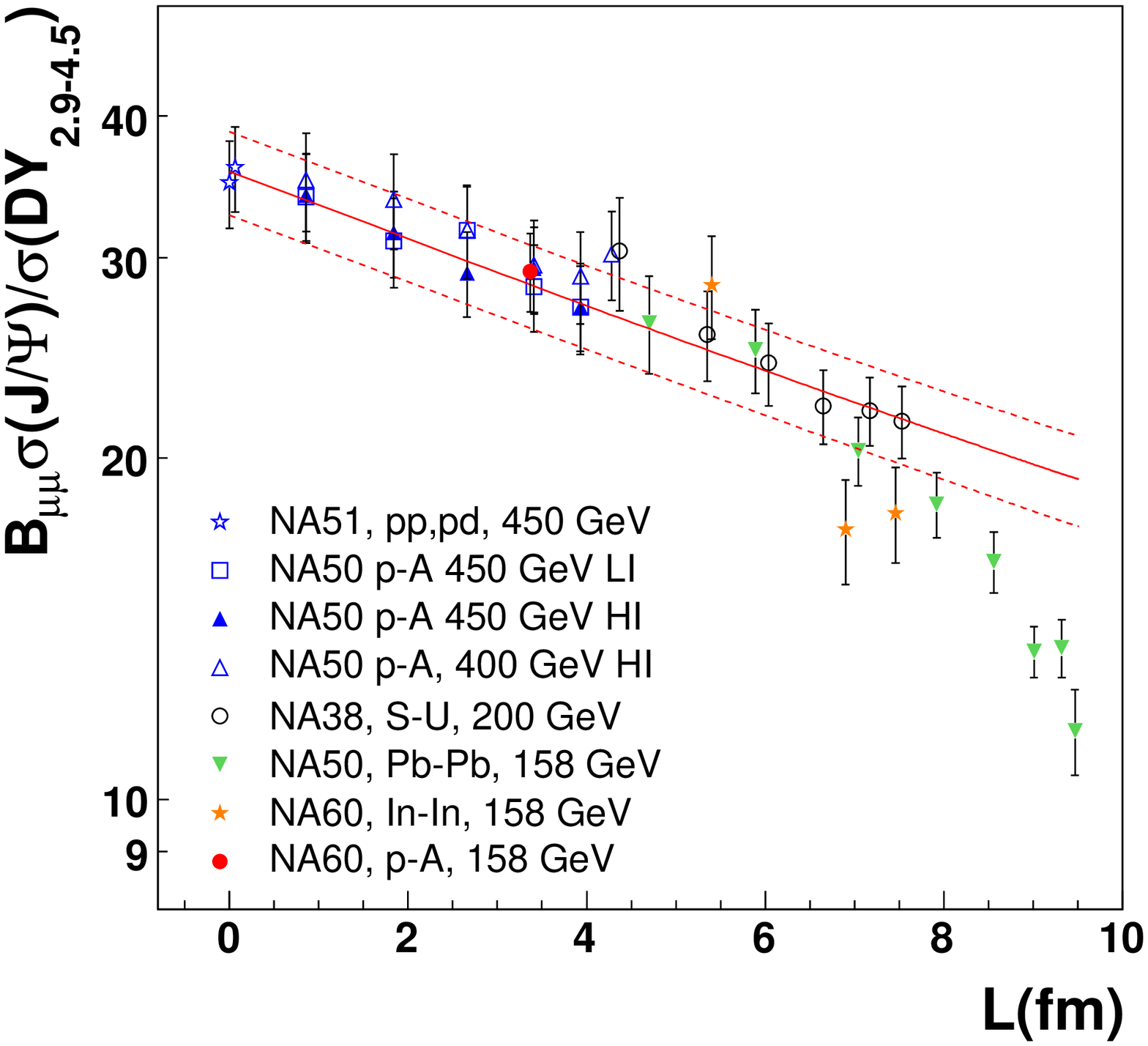} & \includegraphics[width=7.5cm]{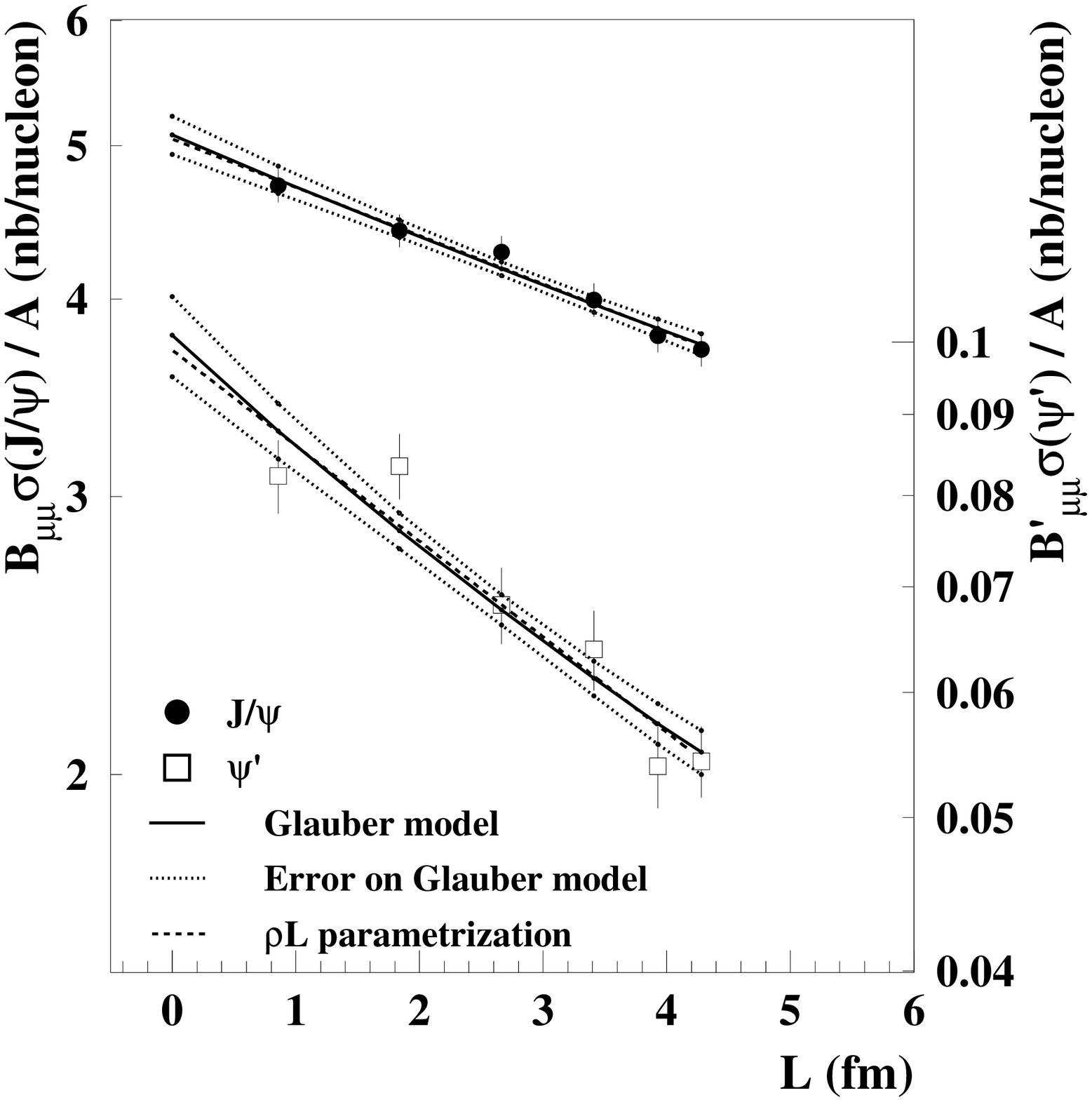}
  \end{tabular}
\end{center}
\vspace{-3ex} \caption{Left: $J/\psi$ yields normalized by Drell-Yan, as a function of the nuclear thickness $L$, as measured at the SPS. Right: $J/\psi$ and $\psi'$ yields in p+A collisions, normalised by A, and fitted to different absorption cross-sections.
} \label{Fig1}
\vspace{-2ex}
\end{figure}

The left part of figure~\ref{Fig1} summarises the $J/\psi$ over Drell-Yan\footnote{As expected, Drell-Yan is not very affected by the medium, as it was reported by NA50~\cite{Abreu:1999qw} and can also be seen on the right part of figure 2 from E722~\cite{Alde:1990im}.} productions in a large variety of collisions, as measured at the CERN SPS ($\sqrt{s_{NN}}\simeq 20$~GeV)~\cite{Scomparin:2007rt}. From \mbox{p+p}, to various p+A, to S+U, and up to peripheral In+In or Pb+Pb collisions, the data falls exponentially as a function of the nuclear thickness parameter $L$. The simple interpretation of this scaling is that $J/\psi$ are absorbed by the forthcoming nucleons, the number of which being proportional to $L$\footnote{Indeed, $L$ is computed by counting the average number $N$ of subsequent collisions in a Glauber model, converting it in a distance through $L=N/\sigma\rho$, where $\rho$ is the nucleon density and $\sigma$ the inelastic N+N cross section (which is finally arbitrary, cancelling out in the $N/\sigma$ ratio).}. From simultaneously taken p+A data, the NA50 experiment extract an absorption cross-section of 4.3~$\pm$~0.5~mb for the $J/\psi$, as well as 7.7~$\pm$~0.9~mb for the higher $\psi'$ state, as seen on the right part of figure~\ref{Fig1}~\cite{Alessandro:2006jt}. As expected, a lesser bound state is more fragile and suffers larger absorption.

Beyond this normal absorption, $J/\psi$ are further suppressed in both the In+In~\cite{Arnaldi:2007zz} and Pb+Pb~\cite{Alessandro:2004ap} more central collisions (right-most points in figure~\ref{Fig1} left), while the $\psi'$ already melts in S+U collisions~\cite{Alessandro:2006jt}.

\section{At FNAL and HERA-B, more quarkonia and $x_F$ scaling}

\begin{figure}[htb]
\begin{center}
  \begin{tabular}{cc}
  \includegraphics[width=6.2cm]{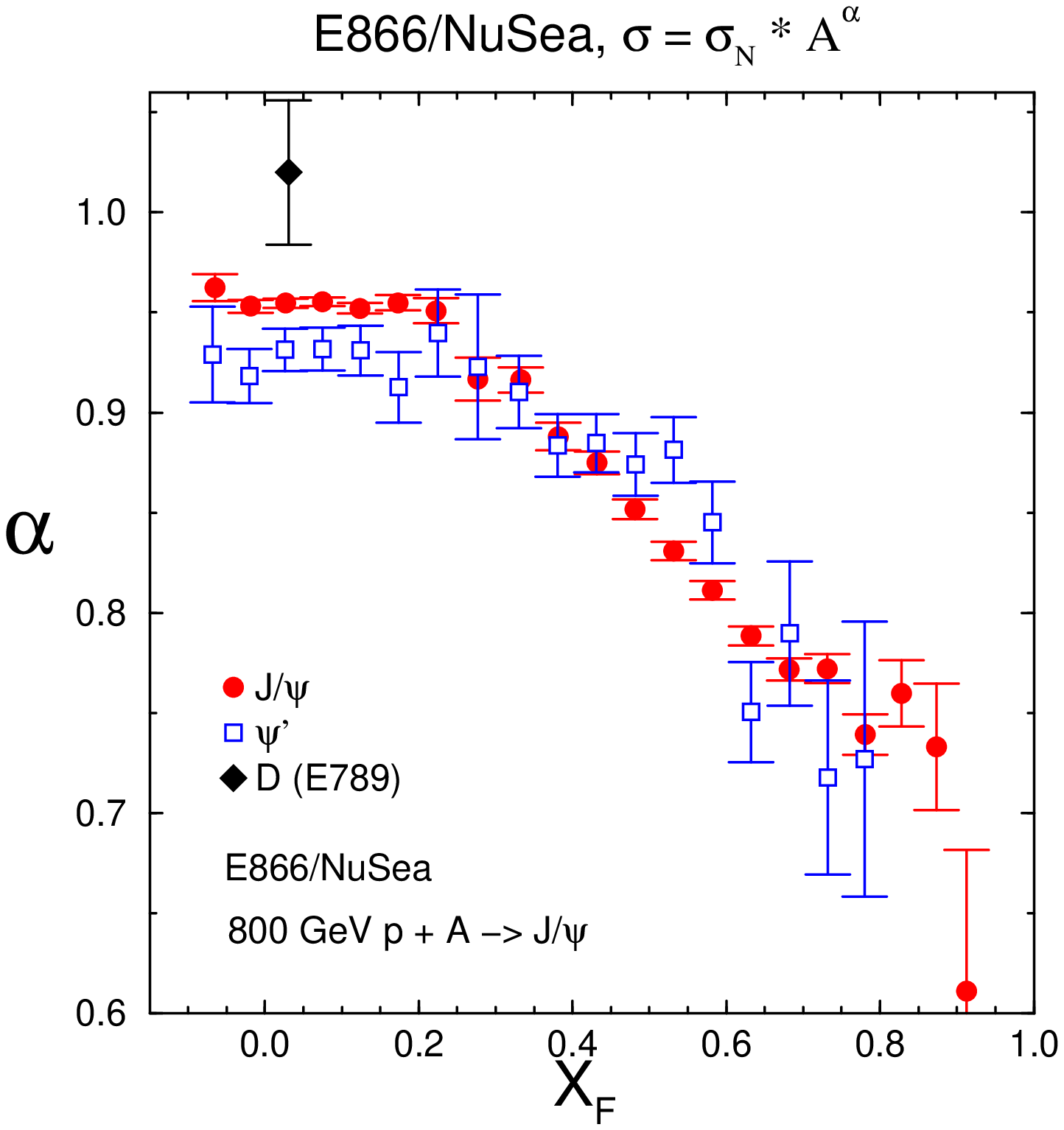} & \includegraphics[width=7.4cm]{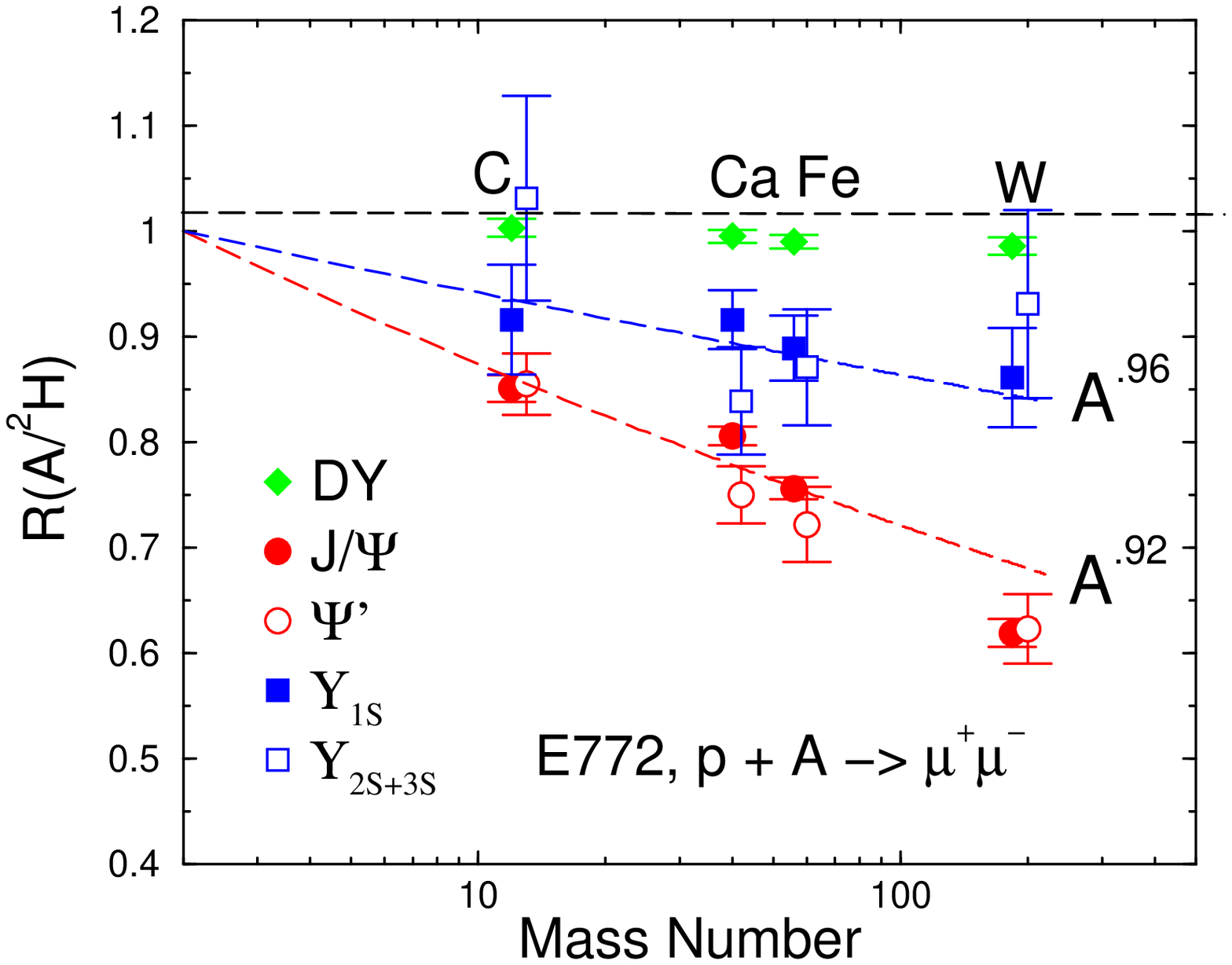}
  \end{tabular}
\end{center}
\vspace{-3ex} \caption{Quarkonia suppression measured at FNAL. Left: through the $\alpha$ parameter versus $x_F$. Right: through the ratio of p+A to p+d cross-sections versus A.
} \label{Fig2}
\vspace{-2ex}
\end{figure}

At slightly higher energy ($\sqrt{s_{NN}}\simeq 40$~GeV), the E866~\cite{Leitch:1999ea} and HERA-B~\cite{Abt:2006va} experiments have also measured that $\psi'$ were more absorbed than $J/\psi$. In the E866 case, this can be seen on the left part of figure~\ref{Fig2} where the power $\alpha$ by which the production is modified with respect to binary scaling ($\sigma_{pA} = \sigma_{pp} \times A^\alpha$) is depicted as a function of Feynman's $x_F = x_p - x_A$ ($x$ refers to the momentum fraction of the parton in the proton projectile $x_p$ or target nucleus $x_A$). Focussing on the $x_F \simeq 0$ region, we indeed see that $\psi'$ are more suppressed than $J/\psi$. It is also the case in the negative $x_F$ region (down to $-0.35$), probed by HERA-B~\cite{Abt:2006va}. In the higher $x_F$ region probed by E866, both the $J/\psi$ and $\psi'$ yields decrease dramatically, as shown on the left part of figure~\ref{Fig2}. This is not explainable without invoking effects beyond simple absorption cross-section, in particular gluon shadowing and energy loss of the incoming parton (see~\cite{Vogt:1999dw} or~\cite{Kopeliovich:2001ee} for two complicated but successful fits of this data). However, since no A+A data exist in this region, we shall focus on $x_F \simeq 0$.

Other particles were significantly measured by these experiments. The upper diamond on the left part of figure~\ref{Fig2} corresponds to D-mesons measured in E789~\cite{Leitch:1994vc}, which (with an accuracy of $\simeq$~4~\%) do not seem to be modified by the medium ($\alpha \simeq 1$). This  was also confirmed recently by HERA-B~\cite{Abt:2007zg}. The right part of figure~\ref{Fig2} compares quasi-unmodified Drell-Yan~\cite{Alde:1990im} (diamonds) to suppressed charmonia~\cite{Alde:1990wa} (circles) and bottomonia~\cite{Alde:1991sw} (squares) in various p+A (normalised to p+d) collisions. Bottomonia suffer less than charmonia from the medium, indicating a smaller (effective) absorption cross-section. HERA-B recently released a detailed preprint~\cite{Abt:2008ed} on $\chi_c$ production, in which the ratio of $J/\psi$ coming from $\chi_c \rightarrow \gamma J/\psi$ is measured to be $18.8 \pm 1.3 \pm 2.4$~\%, not depending on the two (C and W) targets used for this measurement.

\section{A word on a possible $\sigma_{abs}$ energy dependence}

Having measured $J/\psi$ production at several energies, allows one to ask the question: does $\sigma_{abs}$ vary with energy? While theoretical arguments exist to support the hypothesis that it decreases with energy, for instance because the Lorentz boosted and contracted nucleons are already gone when quarkonia form, no firm experimental conclusion exists yet. The interplay between $\sigma_{abs}$ and (at least) shadowing, as well as the different $x$ ranges probed by experiments make it difficult to perform such comparisons, as shown for instance in~\cite{Arleo:2006qk}.

\section{At RHIC, $J/\psi$ and their rapidity dependence}

\begin{figure}[htb]
\begin{center}
  \begin{tabular}{cc}
  \includegraphics[width=7.0cm]{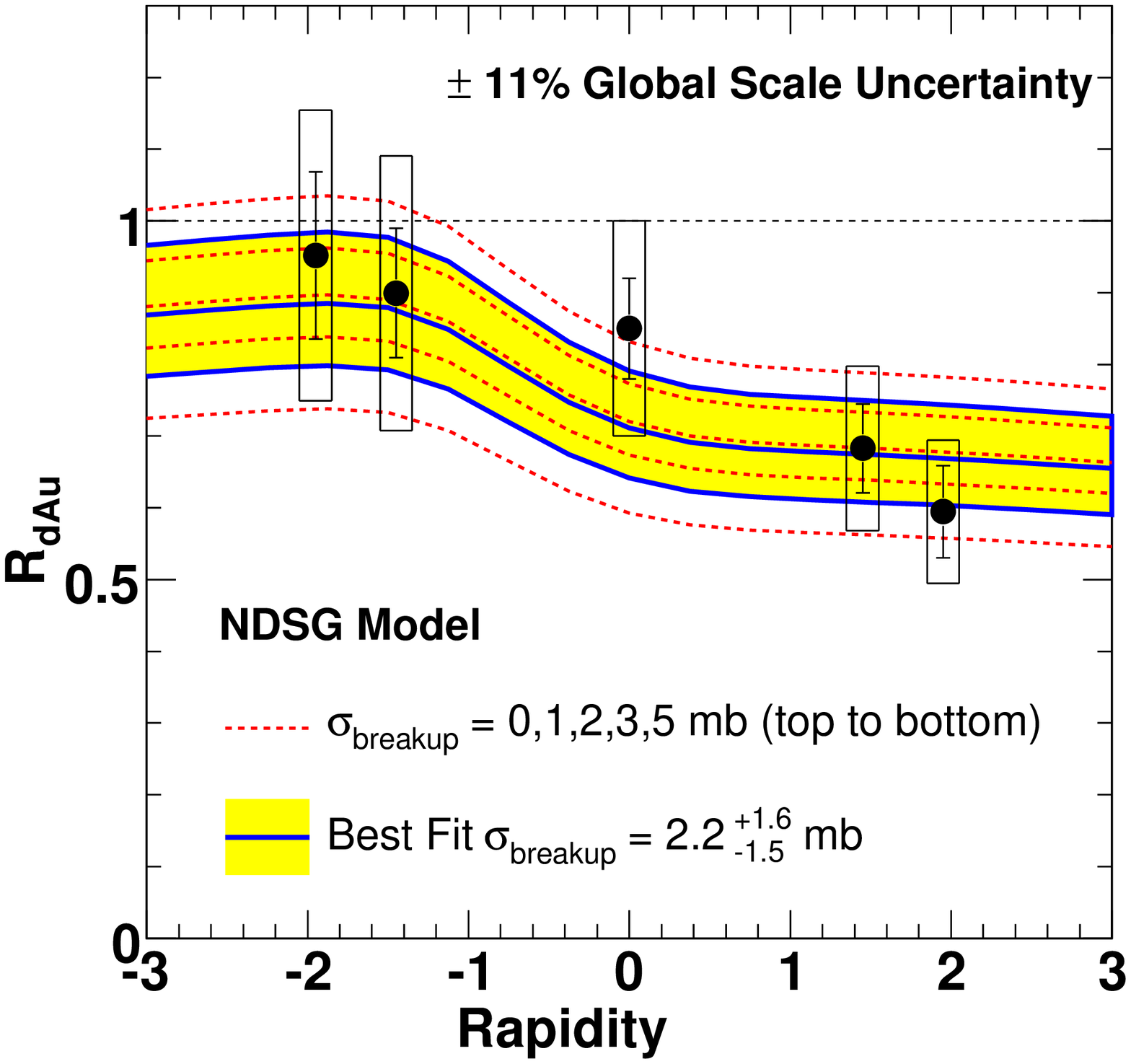} & \includegraphics[width=8.5cm]{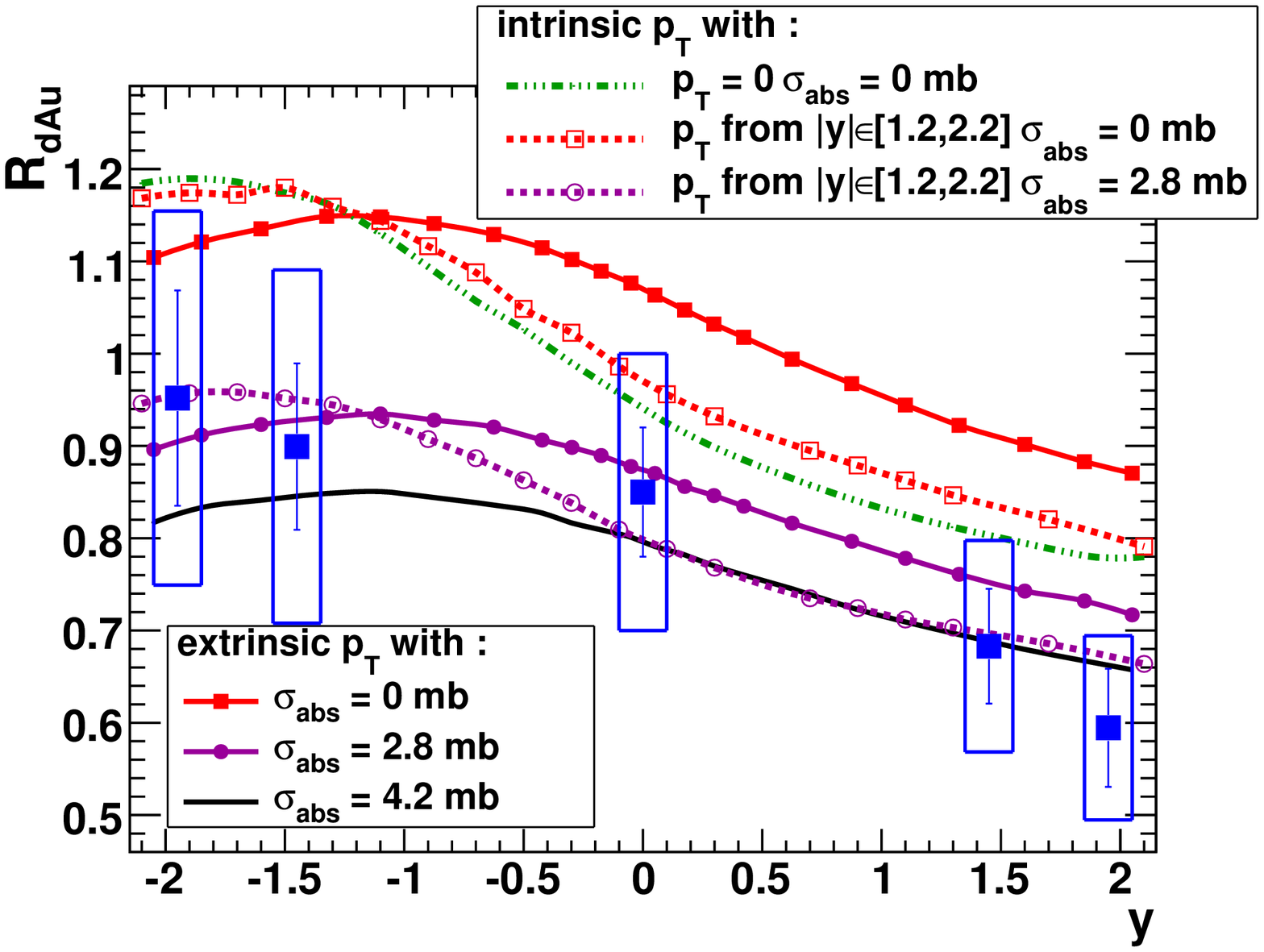}
  \end{tabular}
\end{center}
\vspace{-3ex} \caption{$J/\psi$ nuclear modification factor in d+Au collisions as a function of rapidity, as measured by the PHENIX experiment. Left: Extraction of a break-up cross-section, assuming the NDSG scheme. Right: Illustration of the impact of intrinsic vs. extrinsic production mechanisms.
} \label{Fig3}
\vspace{-2ex}
\end{figure}

At RHIC, the PHENIX experiment is able to measure $J/\psi$ at mid rapidity ($|y|<0.35$), but also at forward rapidity ($1.2<|y|<2.2$), probing a large rapidity range. The left part of figure \ref{Fig3} shows the $J/\psi$ nuclear modification factor measured in \mbox{d+Au} collisions~\cite{Adare:2007gn}. Positive rapidity $J/\psi$ are originating from lower $x$ partons in the Au nuclei (down to $2.10^{-3}$ in this case). They suffer more suppression, and this is usually interpreted as a sign of initial parton shadowing. At this energy ($\sqrt{s_{NN}}=200$~GeV), $c\bar{c}$~pairs mostly originate from the fusion of gluons, the shadowing of which is not precisely constrained by existing data. Assuming various shadowing schemes, one can then derive the additional effect of nuclear absorption. As an example, on the left part of figure~\ref{Fig3} assuming the NDSG~\cite{deFlorian:2003qf} shadowing, a $\sigma_{abs}$ (called $\sigma_{breakup}$ here) value is extracted and quoted on the figure. The experiment confessed at this conference~\cite{LindenLevy:SQM08} that part of the systematic uncertainties was not properly propagated to $\sigma_{abs}$. However, one can already see that cold nuclear effects at RHIC are not well constrained, the uncertainty on $\sigma_{abs}$ being \emph{at least} of 2~mb.

More complications even arise from the underlying mechanism of $J/\psi$ production.
Creating $J/\psi$ quasi-alone (inheriting its $p_T$ from the initial partons \emph{intrinsic} $p_T$) or with a hard gluon (balancing the $J/\psi$ \emph{extrinsic} $p_T$)
should \emph{a priori} probe different $x$ values of the initial partons. Such an effect is derived from the EKS shadowing scheme~\cite{Eskola:1998df} and compared to PHENIX data in a recent preprint~\cite{Ferreiro:2008wc}. Comparing the red open and closed symbols on figure~\ref{Fig3} (right), which differ only by their intrinsic and extrinsic nature, the authors show that such an effect could change $R_{dAu}$ by about 10~\%. This comes on top of a large uncertainty due to the shadowing scheme used.

While it seems that nuclear absorption and shadowing are needed to explain $J/\psi$ production in d+Au collisions at RHIC, it is also clear that there is an interplay between the two, as it is for instance illustrated in~\cite{Arleo:2006qk}, in which the authors derive a 1.7~mb systematic uncertainty on $\sigma_{abs}$ by varying the assumed shadowing scheme. It is also clear that gluon shadowing is poorly constrained by data, as the authors of the EKS model stress in their updated analysis leading to the new EPS scheme~\cite{Eskola:2008ca}.

Two approaches are proposed to deal with this interplay and uncertainties. Trying to avoid relying on (shadowing) models will be addressed in the next section through a data driven method, while the status of open charm measurements, that would allow one to disentangle initial from final state effects will be shown afterwards.

\section{At RHIC, $J/\psi$ and their centrality dependence}

\begin{figure}[htb]
\begin{center}
  \begin{tabular}{cc}
  \includegraphics[width=7.0cm]{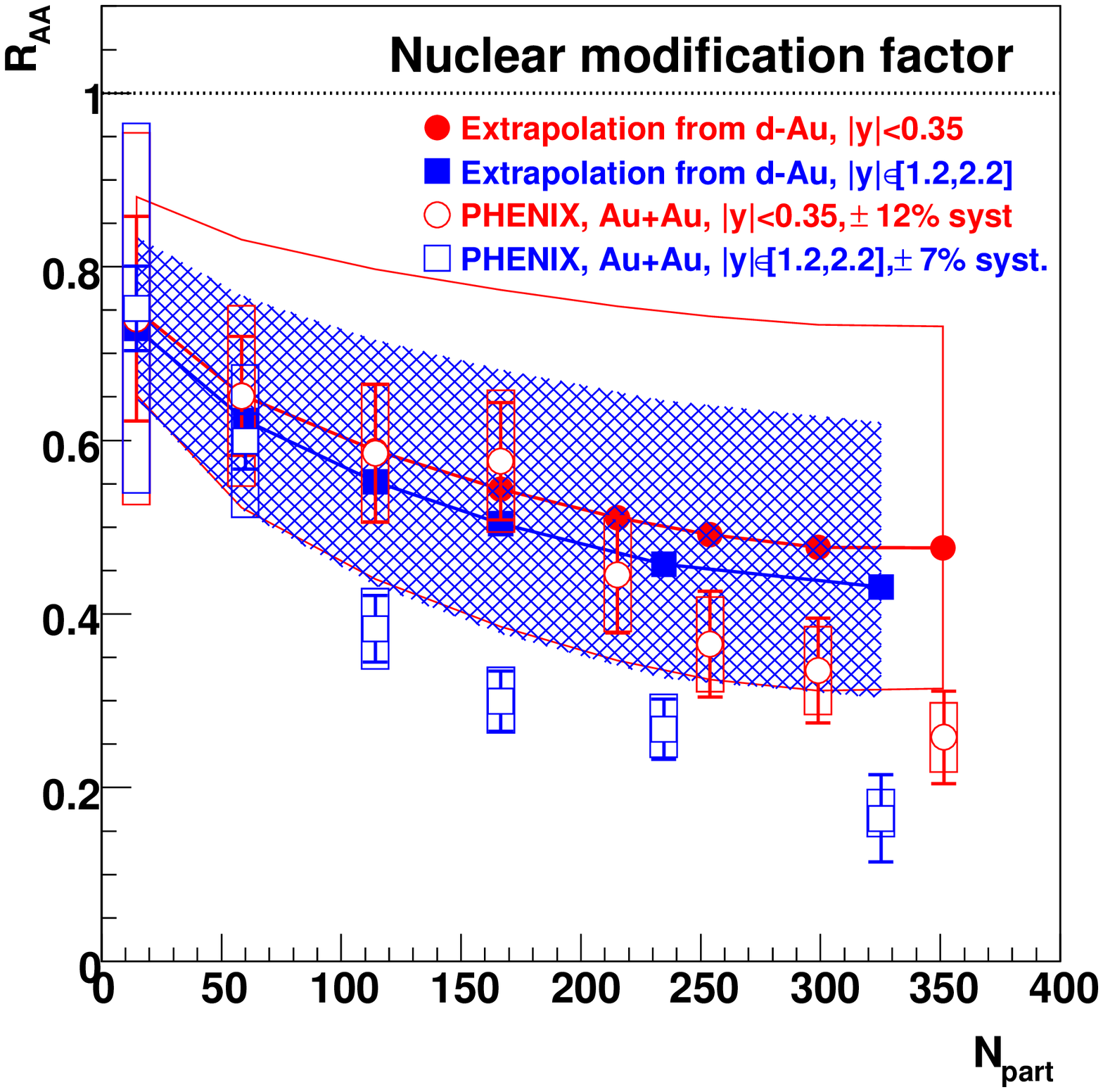} & \includegraphics[width=7.0cm]{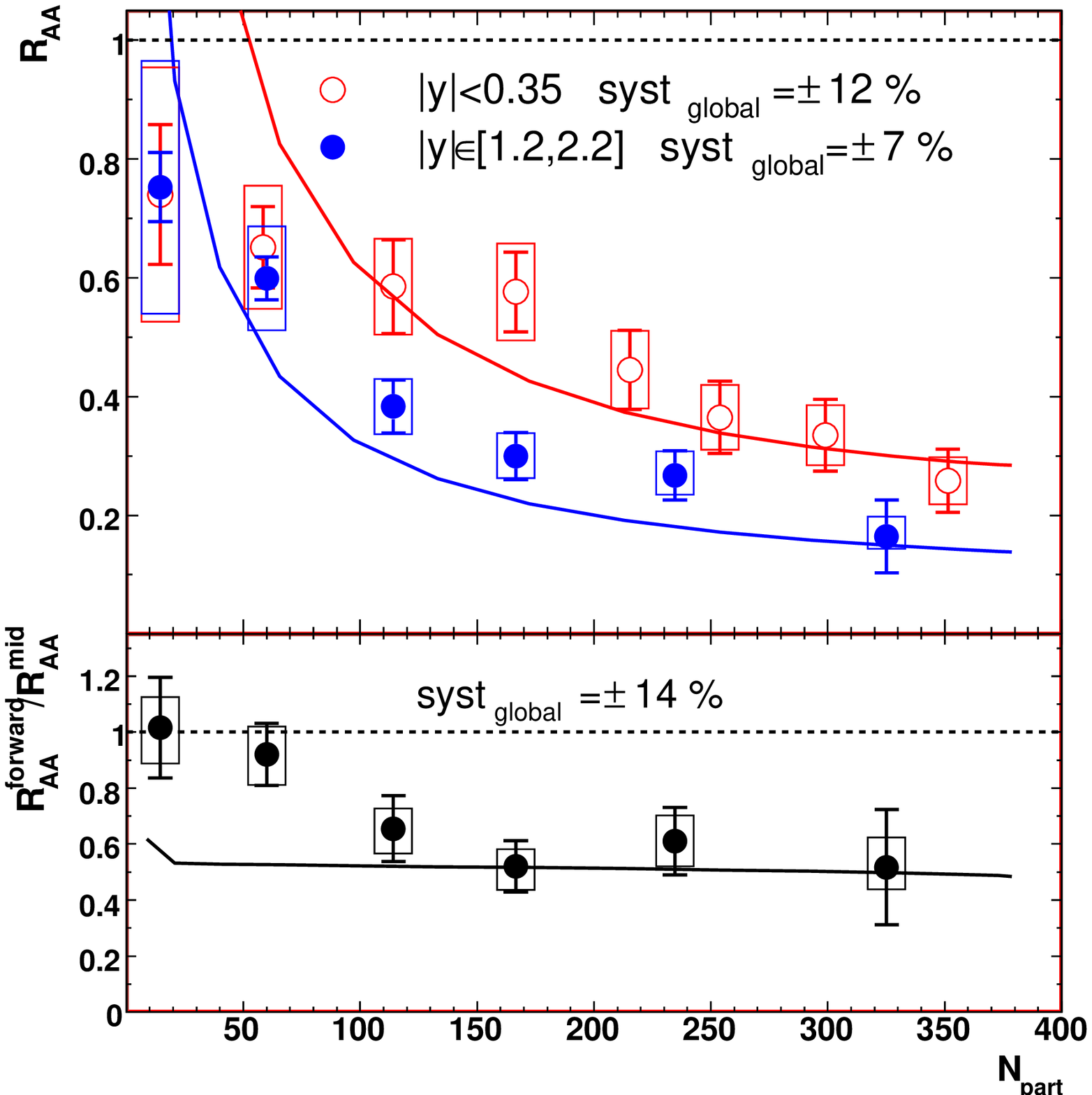}
  \end{tabular}
\end{center}
\vspace{-3ex} \caption{$J/\psi$ $R_{AuAu}$ for the two rapidity measured by PHENIX, as a function of $N_{part}$, and compared to, left: an extrapolation of cold effects from the d+Au measured centrality dependence and to, right: a Colour Glass Condensate calculation.
} \label{Fig4}
\vspace{-2ex}
\end{figure}

Unlike for lower energy p+A like collisions, d+Au centrality is measured at RHIC, and $J/\psi$ have been split in four centrality bins, at the three PHENIX rapidities. In~\cite{deCassagnac:2007aj}, it was proposed to take advantage of this impact parameter dependence to avoid relying on shadowing scheme. To do so, the author first performs phenomenological fits to $R_{dAu}(y,b)$, where $b$ is the impact parameter derived from a Glauber model, also used to extrapolate to Au+Au. For each Au+Au collision occurring at a given impact parameter $b_{AuAu}$, the $N_{coll}$ elementary nucleon-nucleon collisions are randomly distributed (following Woods-Saxon nuclear densities) providing the locations $b^i_1$ and $b^i_2$ of each collision $i$, relative to the centre of nucleus~1 and nucleus~2. For the considered Au+Au collision, the predicted nuclear modification factor $R_{AuAu}$ is given by the following summation over the elementary collisions:
$R_{AuAu} (y, b_{AuAu}) = \sum_{i=1}^{N_{coll}} R_{dAu}(-y,b^i_1) \times R_{dAu}(+y,b^i_2) / N_{coll}$.
 This method has the advantage not to depend on shadowing scheme, $\sigma_{abs}$ or production mechanisms, and to allow an easy propagation of the (large) $R_{dAu}$ experimental uncertainties to $R_{AuAu}$. The latest results of this calculation (from~\cite{Adare:2007gn}) are shown on the left part of figure~\ref{Fig4} and compared to measured $R_{AuAu}$. The obtained authorised cold suppression (depicted as areas) have large enough uncertainties to allow the anomalous suppression to be equal at mid and forward rapidity.

It is to be noted that this conclusion is also reached through the first attempt to derive $J/\psi$ production in the framework of the Colour Glass Condensate (CGC), in a recent preprint~\cite{Kharzeev:2008cv}. While the authors fit the absolute amount of $J/\psi$ suppression to the data itself, its rapidity dependence does not depend on this global fit. The right part of figure~\ref{Fig4} shows a comparison of their result with the measured $R_{AuAu}$. The bottom panel is the ratio $R_{AuAu}(y=1.7)/R_{AuAu}(y=0)$. While this CGC approach manages to reproduce the difference in central collisions, it fails in the more peripheral ones.

\section{At RHIC, a poorly known open charm}

An interesting experimental way that could help disentangling $J/\psi$ nuclear absorption and gluon shadowing would be to measure open charm with good precision. Open charm is indeed sharing its initial state effects (shadowing) with charmonia, while the final state effects should be totally different. At the moment, open charm is known to scale with the number of binary collisions, as we have seen on the left part of figure~\ref{Fig2} for FNAL energies, and is shown in~\cite{Adare:2006nq} for RHIC energies. The latest suffer a systematic uncertainty of the order of 25~\% that can only be reduced with the help of silicon tracking detectors allowing to measure displaced vertices.

\begin{figure}[htb]
\begin{center}
  \begin{tabular}{cc}
  \includegraphics[width=6.2cm]{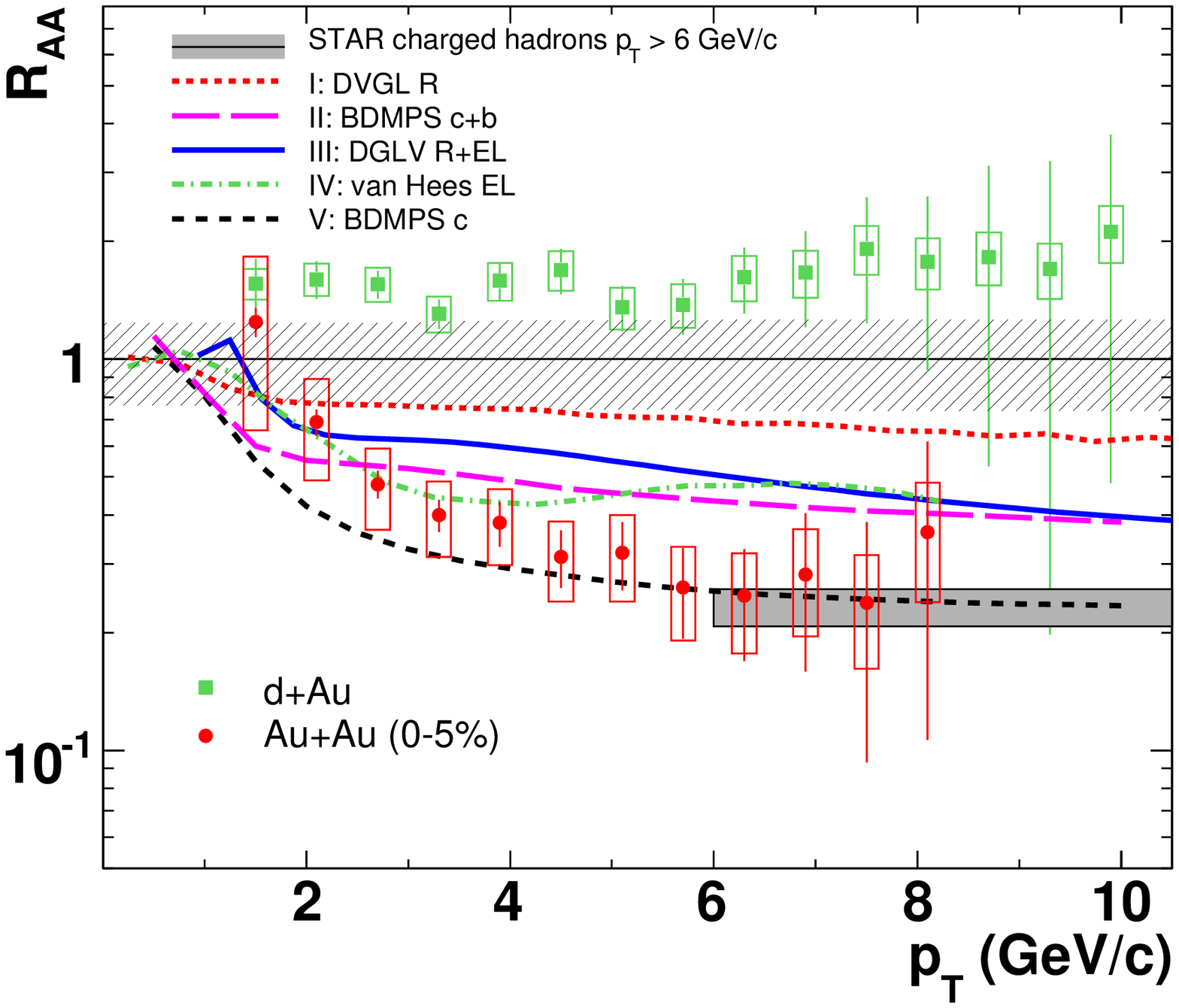} & \includegraphics[width=7.4cm]{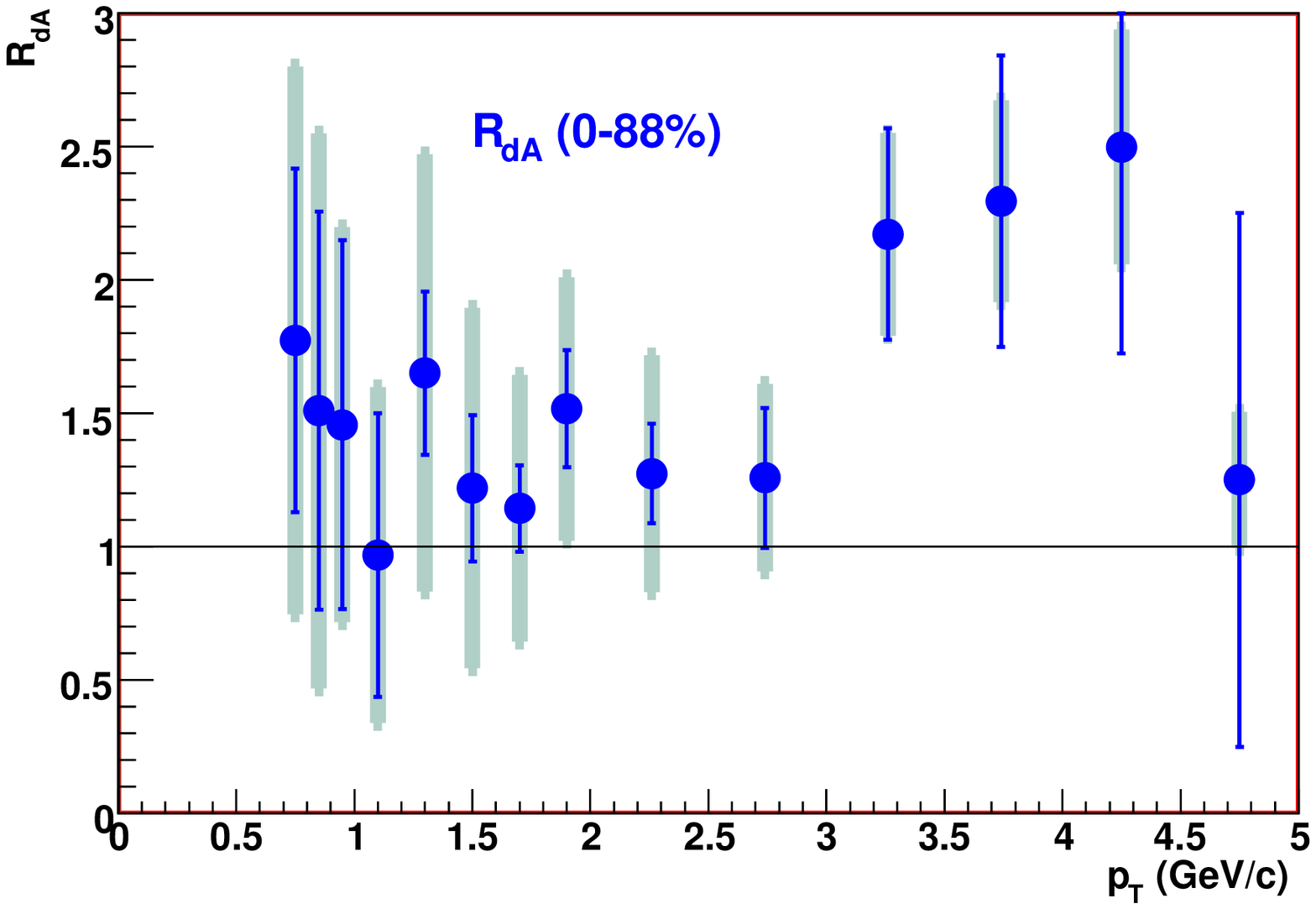}
  \end{tabular}
\end{center}
\vspace{-3ex} \caption{Nuclear modification factor of electrons from heavy flavour decays. Left: from STAR. Right: computed from PHENIX preliminary results.
} \label{Fig5}
\vspace{-2ex}
\end{figure}

While the bulk of heavy flavour production scales with binary collisions, both PHENIX~\cite{Adare:2006nq} and STAR~\cite{Abelev:2006db} have reported a large suppression of electrons coming from heavy flavour decays at high $p_T$. The situation in d+Au collision is less clear, as it can be seen on figure~\ref{Fig5}. The left part is the STAR measurement of the Au+Au (red circles) and d+Au (green squares) nuclear modification factors~\cite{Abelev:2006db}, while the right part reflects an extraction of the same quantity I made, based on preliminary d+Au~\cite{Kelly:2004qw} and published p+p~\cite{Adare:2006hc} PHENIX data. Both figures exhibit an interesting enhancement that will be much better addressed by the analysis of the 2008 RHIC data which have 30~times the statistics used here. If confirmed, such an enhancement could be interpreted in terms of antishadowing and/or Cronin effect, depending on its $p_T$ behaviour.

\section{And the Cronin effect everywhere}

\begin{figure}[htb]
\begin{center}
  \begin{tabular}{cc}
%   \includegraphics[height=5.4cm]{Fig6_NA60.eps}
% & \hspace{-3em} \includegraphics[height=5.2cm]{Fig6_PHENIX.eps}
  \includegraphics[height=5.4cm]{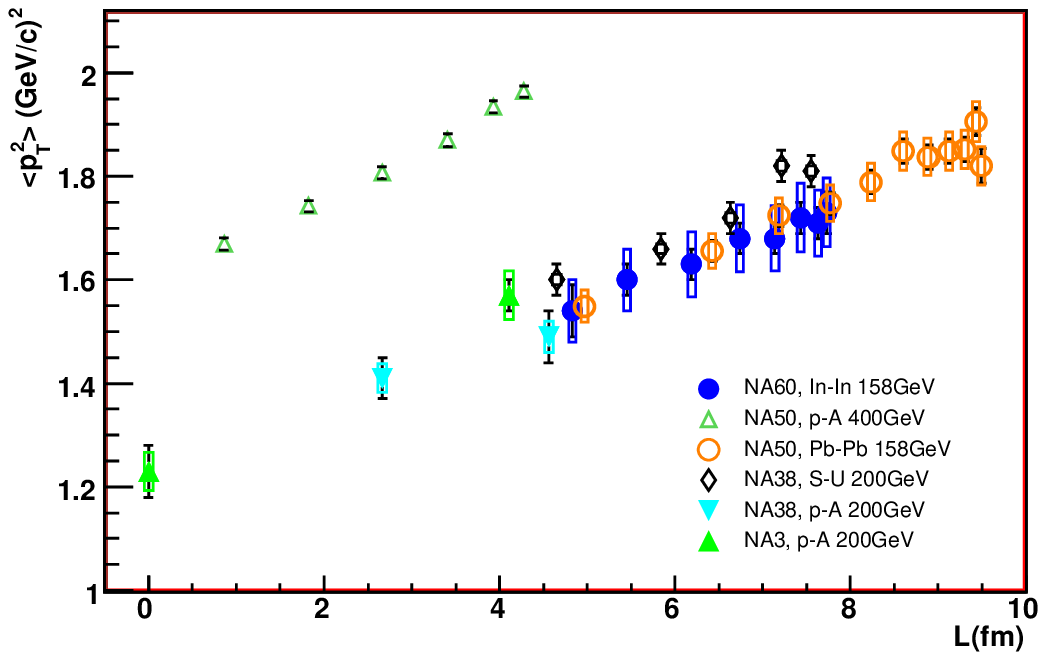}
 & \hspace{-3em} \includegraphics[height=5.2cm]{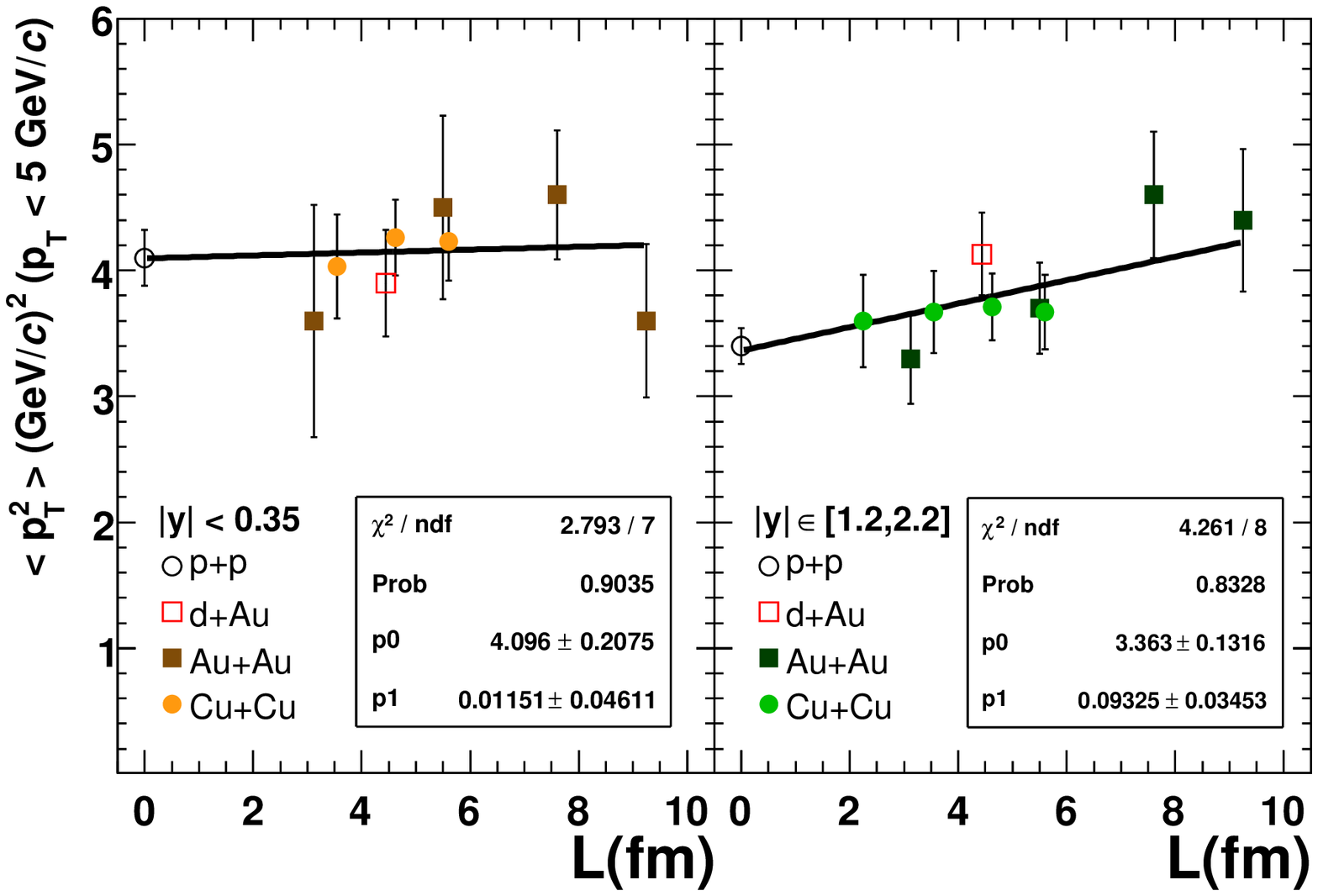}
  \end{tabular}
\end{center}
\vspace{-2ex} \caption{$J/\psi$ $<p_T^2>$ versus $L$ from (left) SPS experiments (right) PHENIX.
} \label{Fig6}
\vspace{-2ex}
\end{figure}

The so-called Cronin effect consists in multiple scattering of an initial parton on the facing nucleus, resulting in an increased $p_T$ of the final state. In E866 for instance, a raise of $\alpha$ with $p_T$ is observed for Drell-Yan, charmonia~\cite{Leitch:1999ea} and bottomonia~\cite{Alde:1991sw}. The average resulting $p_T^2$ should vary proportionally to the amount of centres the partons can scatter upon, that can be characterised by $L$, the thickness parameter\footnote{Indeed, for instantaneous processes, the average number of subsequent or preceding interactions are proportional, and the same $L$ thickness parameter can be considered for initial (Cronin) of final (nuclear absorption) effects.}. The left part of figure~\ref{Fig6} shows $<p_T^2>$ as a function of $L$, as measured by SPS experiments\footnote{The missing NA60 p+A measurement was shown in conferences but not published.}. For each energy, it clearly exhibits the expected linear dependence. The left part of figure~\ref{Fig6} shows the current status of $<p_T^2>$ measurements at RHIC~\cite{Adare:2008sh}. I plot them here as a function of $L$ and perform simple linear fits to all the points, including p+p, d+Au, Cu+Cu and Au+Au data. Within this limited accuracy, the mid rapidity slope is compatible with zero while the forward slope has a significance of 2.7~standard deviations, which could be a first sign of Cronin effect. This measurement clearly deserves more precision which should come soon for d+Au (2008 run) and Au+Au (2007 run).

\section{LHC, the uncharted territory}

As a conclusion, let us think about the future LHC energy regime. In the previous sections, we saw that cold effects affecting heavy flavours were not well understood when one considers a large amount of data (large $x_F$ coverage or various $\sqrt{s_{NN}}$). They are also not well constrained by experimental data, especially at RHIC. It would thus be very daring to predict them at LHC. As an exemple, $J/\psi$ at midrapidity will be produced from the fusion of two gluons with $x \lesssim 10^{-3}$. The difference between various shadowing models in this regime is about a factor of two (to be squared in the fusion), for $Q^2\simeq m_c^2$. It will thus be crucial to have p+A like collisions at LHC, in order to interpret heavy flavour productions...

\section*{Acknowledgments}

The author would like to thank Roberta Arnaldi and Mike Leitch for fruitful discussions and providing some of the figures shown here.

\section*{References}

\bibliographystyle{myunsrt}
\bibliography{Biblio}

\begin{thebibliography}{10}

\bibitem{Adare:2006ns}
A.~Adare et~al.
\newblock Phys. Rev. Lett. 98 (2007) 232201, nucl-ex/0611020.

\bibitem{Matsui:1986dk}
T.~Matsui and H.~Satz.
\newblock Phys. Lett. B178 (1986) 416.

\bibitem{LindenLevy:SQM08}
L.~A. Linden-Levy (PHENIX~collaboration).
\newblock This conference.

\bibitem{Gunji:SQM08}
T.~Gunji.
\newblock This conference.

\bibitem{Abreu:1999qw}
M.~C. Abreu et~al.
\newblock Phys. Lett. B450 (1999) 456-466.

\bibitem{Alde:1990im}
D.~M. Alde et~al.
\newblock Phys. Rev. Lett. 64 (1990) 2479-2482.

\bibitem{Scomparin:2007rt}
E.~Scomparin.
\newblock J. Phys. G34 (2007) S463-470, nucl-ex/0703030.

\bibitem{Alessandro:2006jt}
B.~Alessandro et~al.
\newblock Eur. Phys. J. C48 (2006) 329, nucl-ex/0612012.

\bibitem{Arnaldi:2007zz}
R.~Arnaldi et~al.
\newblock Phys. Rev. Lett. 99 (2007) 132302.

\bibitem{Alessandro:2004ap}
B.~Alessandro et~al.
\newblock Eur. Phys. J. C39 (2005) 335, hep-ex/0412036.

\bibitem{Leitch:1999ea}
M.~J. Leitch et~al.
\newblock Phys. Rev. Lett. 84 (2000) 3256-3260, nucl-ex/9909007.

\bibitem{Abt:2006va}
I.~Abt et~al.
\newblock Eur. Phys. J. C49 (2007) 545-558, hep-ex/0607046.

\bibitem{Vogt:1999dw}
R.~Vogt.
\newblock Phys. Rev. C61 (2000) 035203, hep-ph/9907317.

\bibitem{Kopeliovich:2001ee}
B.~Kopeliovich, A.~Tarasov, and J.~Hufner.
\newblock Nucl. Phys. A696 (2001) 669-714, hep-ph/0104256.

\bibitem{Leitch:1994vc}
M.~J. Leitch et~al.
\newblock Phys. Rev. Lett. 72 (1994) 2542-2545.

\bibitem{Abt:2007zg}
I.~Abt et~al.
\newblock Eur. Phys. J. C52 (2007) 531-542, 0708.1443.

\bibitem{Alde:1990wa}
D.~M. Alde et~al.
\newblock Phys. Rev. Lett. 66 (1991) 133-136.

\bibitem{Alde:1991sw}
D.~M. Alde et~al.
\newblock Phys. Rev. Lett. 66 (1991) 2285-2288.

\bibitem{Abt:2008ed}
I.~Abt et~al.
\newblock (HERA-B collaboration), arXiv:0807.2167.

\bibitem{Arleo:2006qk}
F.~Arleo and V.-N. Tram.
\newblock Eur. Phys. J. C55 (2008) 449-461, hep-ph/0612043.

\bibitem{Adare:2007gn}
A.~Adare et~al.
\newblock Phys. Rev. C77 (2008) 024912, arXiv:0711.3917.

\bibitem{deFlorian:2003qf}
D.~de~Florian and R.~Sassot.
\newblock Phys. Rev. D69 (2004) 074028, hep-ph/0311227.

\bibitem{Eskola:1998df}
K.~J. Eskola, V.~J. Kolhinen, and C.~A. Salgado.
\newblock Eur. Phys. J. C9 (1999) 61-68.

\bibitem{Ferreiro:2008wc}
E.~G. Ferreiro, F.~Fleuret, J.~P. Lansberg, and A.~Rakotozafindrabe.
\newblock arXiv:0809.4684.

\bibitem{Eskola:2008ca}
K.~J. Eskola, H.~Paukkunen, and C.~A. Salgado.
\newblock JHEP 07 (2008) 102, 0802.0139.

\bibitem{deCassagnac:2007aj}
R.~Granier~de Cassagnac.
\newblock J. Phys. G34 (2007) S955, hep-ph/0701222.

\bibitem{Kharzeev:2008cv}
D.~Kharzeev, E.~Levin, M.~Nardi, and K.~Tuchin.
\newblock arXiv:0809.2933.

\bibitem{Adare:2006nq}
A.~Adare et~al.
\newblock Phys. Rev. Lett. 98 (2007) 172301, nucl-ex/0611018.

\bibitem{Abelev:2006db}
B.~I. Abelev et~al.
\newblock Phys. Rev. Lett. 98 (2007) 192301, nucl-ex/0607012.

\bibitem{Kelly:2004qw}
S.~Kelly.
\newblock J. Phys. G30 (2004) S1189-S1192, nucl-ex/0403057.

\bibitem{Adare:2006hc}
A.~Adare et~al.
\newblock Phys. Rev. Lett. 97 (2006) 252002, hep-ex/0609010.

\bibitem{Adare:2008sh}
A.~Adare et~al.
\newblock Phys. Rev. Lett. 101 (2008) 122301, arXiv:0801.0220.

\end{thebibliography}

%\begin{thebibliography}{10}
%\bibitem{book1} Goosens M, Rahtz S and Mittelbach F 1997 {\it The \LaTeX\ Graphics Companion\/}
%(Reading, MA: Addison-Wesley)
%\bibitem{eps} Reckdahl K 1997 {\it Using Imported Graphics in \LaTeX\ } (search CTAN for the file `epslatex.pdf')
%\end{thebibliography}

\end{document}